\documentclass{article}

\PassOptionsToPackage{numbers, compress}{natbib}
 \usepackage[preprint]{neurips_2025}

\usepackage[utf8]{inputenc} 
\usepackage[T1]{fontenc}    
\usepackage{hyperref}       
\usepackage{url}            
\usepackage{booktabs}       
\usepackage{amsfonts}       
\usepackage{nicefrac}       
\usepackage{microtype}      
\usepackage{xcolor}         
\usepackage{tcolorbox}
\tcbuselibrary{skins,breakable,listings}
\newenvironment{cuteexample}[1][]
{\begin{tcolorbox}[title=#1, colback=yellow!10, colframe=orange!50!black, fonttitle=\bfseries, breakable]}
{\end{tcolorbox}}

\usepackage{amsmath} 
\usepackage{booktabs}
\usepackage{float} 
\usepackage{enumitem}

\usepackage{amsthm}
\usepackage{mdframed}

\newmdtheoremenv[
  linecolor=gray!20,
  linewidth=1pt,
  backgroundcolor=gray!10,
  innertopmargin=4pt,
  innerbottommargin=4pt,
]{definition}{Definition}
\newtheorem{remark}{Remark}
\def\I{I}

\title{An Outlook on the Opportunities and Challenges of Multi-Agent AI Systems}

\author{%
Fangqiao Tian\textsuperscript{1},
An Luo\textsuperscript{1},
Jin Du\textsuperscript{1},
Xun Xian\textsuperscript{2},
Robert Specht\textsuperscript{1},
Ganghua Wang\textsuperscript{3},\\
Xuan Bi\textsuperscript{4},
Jiawei Zhou\textsuperscript{5},
Ashish Kundu\textsuperscript{6},
Jayanth Srinivasa\textsuperscript{6},
Charles Fleming\textsuperscript{6},\\
Rui Zhang\textsuperscript{7},
Zirui Liu\textsuperscript{8},
Mingyi Hong\textsuperscript{2},
Jie Ding\textsuperscript{1}
}

\makeatletter
\renewcommand{\@maketitle}{
  \newpage
  \null
  \vskip 1em%
  \begin{center}%
    {\LARGE \bfseries \@title \par}%
    \vskip 1.5em%
    {
    \@author \par
    \vskip 0.5em
    \small
    \begin{center}
    \textsuperscript{1}School of Statistics, University of Minnesota\\
    \textsuperscript{2}Department of Electrical \& Computer Engineering, University of Minnesota\\
    \textsuperscript{3}Data Science Institute, University of Chicago\\
     \textsuperscript{4}Carlson School of Management, University of Minnesota\\
    \textsuperscript{5}Department of Applied Mathematics \& Statistics, Stony Brook University\\
    \textsuperscript{6}Cisco Research\\
    \textsuperscript{7}Department of Surgery, University of Minnesota\\
    \textsuperscript{8}Department of Computer Science \& Engineering, University of Minnesota\\
    
    \end{center}
    \vskip 1.5em
    }
  \end{center}%
}
\makeatother

\begin{document}
\maketitle

\begin{abstract}

A multi-agent AI system (MAS) is composed of multiple autonomous agents that interact, exchange information, and make decisions based on internal generative models. Recent advances in large language models and tool-using agents have made MAS increasingly practical in areas like scientific discovery and collaborative automation. However, key questions remain: When are MAS more effective than single-agent systems? What new safety risks arise from agent interactions? And how should we evaluate their reliability and structure? This paper outlines a formal framework for analyzing MAS, focusing on two core aspects: effectiveness and safety. We explore whether MAS truly improve robustness, adaptability, and performance, or merely repackage known techniques like ensemble learning. We also study how inter-agent dynamics may amplify or suppress system vulnerabilities. While MAS are relatively new to the signal processing community, we envision them as a powerful abstraction that extends classical tools like distributed estimation and sensor fusion to higher-level, policy-driven inference. Through experiments on data science automation, we highlight the potential of MAS to reshape how signal processing systems are designed and trusted.

\end{abstract}

\section{Introduction}

Multi-agent AI systems (MAS) have emerged as a leading paradigm for addressing complex tasks beyond the capabilities of single-agent solutions. Reflecting the growing industrial adoption of MAS, the global AI agents market is expected to surge from roughly \$3.66~billion in 2023 to \$139.12~billion by 2033, with a compound annual growth rate (CAGR) of about 43.88\%~\cite{MarketUS2024}. Industry analyses further indicate MAS could significantly impact the global economy, potentially increasing global GDP by over 25\% within the next decades and establishing MAS as the basis of the next trillion-dollar AI market. Recent open-source frameworks such as \emph{AutoGen}~\cite{AutoGen2023}, \emph{AutoGPT}~\cite{AutoGPT2023}, and \emph{AGNTCY}~\cite{agntcy_website_2025} showcased diverse and successful implementations of agent communication and task assignment. 

\begin{figure}[tb]
\centering
\includegraphics[width=0.85\textwidth]{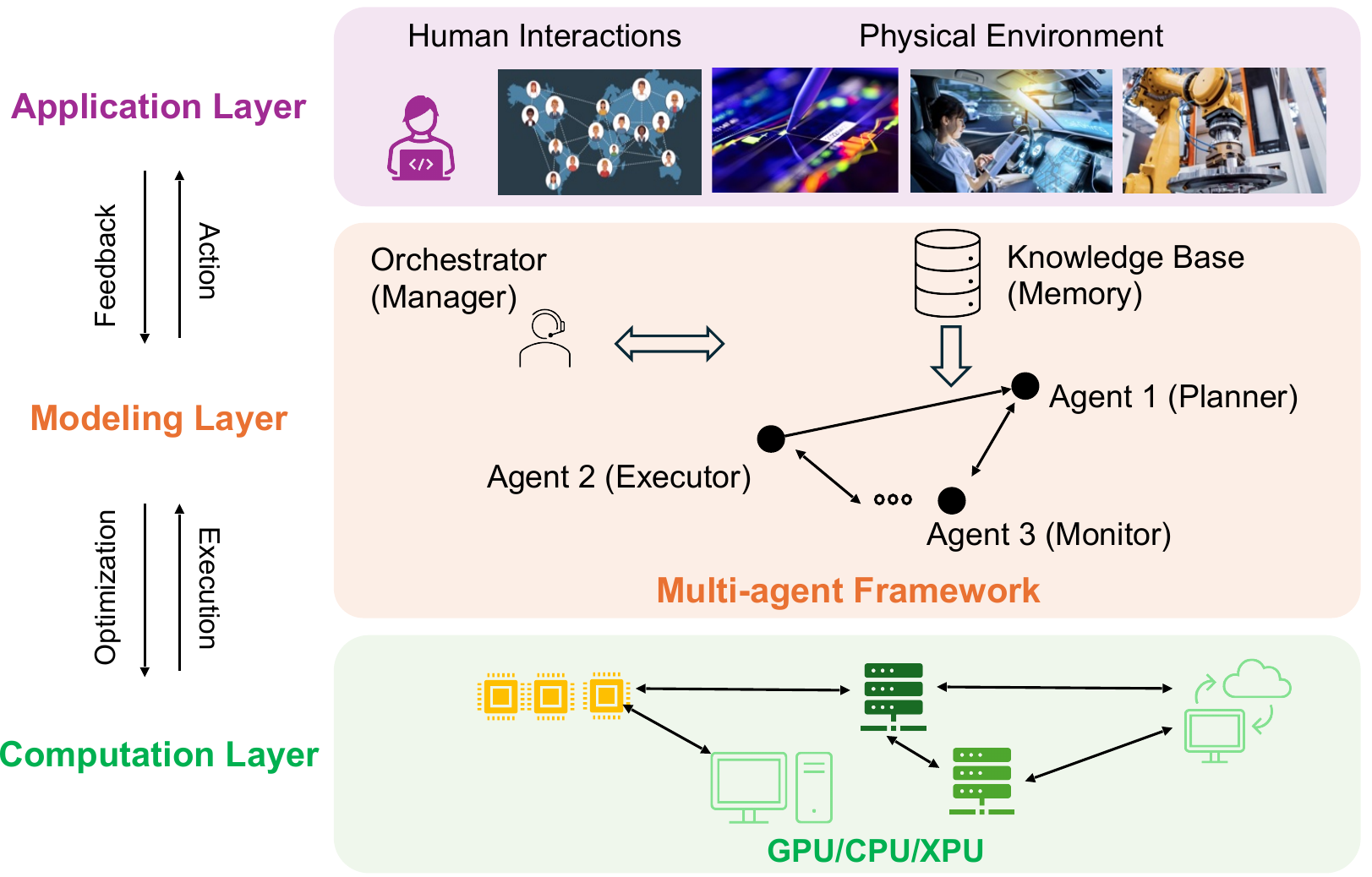}
\caption{Architecture of a multi‑agent AI system with three interconnected layers: (1) Application Layer, interacting dynamically with human users and physical environments through feedback and actions; (2) Modeling Layer, featuring specialized agents (Planner, Executor, and Monitor) coordinated by an orchestrator and  a knowledge base; and (3) Computation Layer, comprising hardware infrastructures (GPU/CPU/XPU) for executing and optimizing agent-driven computations.}
\label{fig:overarching}
\end{figure}

The concept of multi-agent systems has its roots in decades of interdisciplinary research.
Game theory~\cite{Neumann1944} provided formal frameworks to analyze strategic interactions among rational agents. Information theory, driven by seminal works from Shannon~\cite{Shannon1948} and Kolmogorov~\cite{Kolmogorov1965}, provide mechanisms crucial for encoding, transmission, and signal interpretation among agents. Cybernetics~\cite{Wiener1948} highlighted the critical role of feedback loops for adaptive control. Early robotics experiments, such as those conducted by W. Grey Walter~\cite{Walter1950}, demonstrated the potential for autonomous interactions within dynamic environments. Later, machine learning, particularly deep reinforcement learning~\cite{Littman1994,Mnih2015,Foerster2016,Lowe2017}, further advanced these concepts.

Despite the rapid development and deployment of MAS, a formal definition of MAS remains elusive, especially as new agent modalities and interactive structures continually emerge. Prior surveys have extensively covered traditional multi-agent learning and coordination theory~\cite{Shoham2008, Stone2000, Wooldridge2009}. However, they typically focus on the classical MAS perspective, insufficient for modern large-scale needs. Such theories often predate or loosely address today's generative, LLM-based multi-agent AI paradigms. Although some recent overviews have reviewed multi-agent AI systems~\cite{guo2024llm, han2024challenges}, they mostly summarize existing frameworks and benchmarks without explicitly defining what multi-agent actually means, nor contrasting multi-agent and single-agent solutions in a principled manner.  

The resulting conceptual ambiguity around MAS leaves crucial questions regarding their effectiveness and safety unanswered. To bridge this gap, our paper advances the following position: Designing better MAS fundamentally requires understanding under which precise conditions MAS significantly outperform single-agent systems, and how multi-agent setups influence or exacerbate existing safety concerns. We therefore examine these two pivotal dimensions, effectiveness and safety. Our goal is to stimulate principled research towards robust and trustworthy multi-agent AI systems.

\section{Definition and Formulation of Multi-Agent AI Systems}
\label{sec_mas}

We present a formal framework to describe how autonomous agents operate and learn within a shared system. Specifically, Section~\ref{subsec_mas_notation} introduces notations and key concepts of multi-agent AI systems (MAS). Section~\ref{subsec_feedback_notation} discusses how {feedback} can be incorporated to iteratively update MAS. Section~\ref{AGNTCY} further extends this framework to model the internet of agentic systems in open-world, networked environments, inspired by the recently proposed AGNTCY architecture~\cite{cisco_whitepaper_agntcy}.

\subsection{Core Elements}
\label{subsec_mas_notation}

An MAS consists of a set of autonomous agents \( \{i \in \mathcal{I}\} \) that interact through a dynamic communication graph and collectively process inputs over time. Each agent operates via a probabilistic mapping from prior internal state and current input to new state and output. We start by defining the fundamental building block of the MAS: the individual \emph{AI agent}.

\begin{definition}[AI Agent]
\label{def:MAS_agent}
An \emph{AI agent} $i \in \I$ is defined by a tuple $(\mathcal{S}_i, \mathcal{X}_i, \mathcal{Y}_i, p_i)$, where:
\begin{itemize}
    \item $\mathcal{S}_i$ is the agent's internal state space representing its memory or context.
    \item $\mathcal{X}_i$ is the input space defined as a tuple combining outputs from other agents and exogenous feedback:
    \[
        x_i^{(t)} = \left(\left\{y_j^{(t-1)}\mid j \in V^{(t-1)}, (j \rightarrow i) \in E^{(t-1)}\right\}, \mathcal{F}_i^{(t)}\right)
    \]
    where $y_j^{(t-1)}$ denotes outputs from agent $j$ at the previous time step.
    \item $\mathcal{F}_i^{(t)}$ represents exogenous feedback (e.g., signals or human inputs from the physical environment) received by the agent at time $t$.
    \item $\mathcal{Y}_i$ is the output space representing all actions or messages the agent may produce.
    \item $p_i$ is a transition kernel updating the agent’s internal state and output:
    \[
        (s_i^{(t)}, y_i^{(t)}) \sim p_i\left(\cdot \mid s_i^{(t-1)}, x_i^{(t)}\right),
    \]
    where $s_i^{(t)} \in \mathcal{S}_i$ and $y_i^{(t)} \in \mathcal{Y}_i$ are the internal state and output at time $t$, respectively, given the previous state $s_i^{(t-1)}$ and the current input $x_i^{(t)} \in \mathcal{X}_i$.
    \end{itemize}
\end{definition}

\begin{remark}
We track discrete time $t$ so that feedback events can trigger updates to connections and agent states in a systematic manner. In simpler cases, one might omit time dependence to study only static or batch-style multi-agent interactions.
\end{remark}

\begin{remark}
Each agent's input $x_i^{(t)}$ can incorporate outputs from its connected agents in the last round and exogenous feedback, namely $y_j^{t-1}$ for $j \in \I$ and $\mathcal{F}_i^{(t)}$.
\end{remark}

\begin{remark}[On flexibility of input timing]
We define that $x_i^{(t)}$ uses $y_j^{(t-1)}$ as input to ensure that all agents update at time $t$ are causally well-posed and can be computed in parallel. 

However, in systems where a partial or total update order is required (for example, an agent topology structured as DAG), inputs may include some outputs $y_j^{(t)}$ from agent $j$ that are updated strictly before agent $i$ at time $t$. 
In this case, we define
\[
x_i^{(t)} = \left( \left\{ y_j^{(t')} \mid t' \in \{t-1,\ t\},\ j \prec i \text{ if } t'=t \right\},\ \mathcal{F}_i^{(t)} \right)
\]
where the update precedence relation $j \prec i$ is any valid topological order of agent updates.
\end{remark}

Given these individual agents, we next formalize how they interact by defining the concept of dynamic \emph{multi-agent topology}.

\begin{definition}[Multi-Agent Topology]
\label{def:MAS_topology}
An \emph{multi-agent topology} at time $t$ is a directed graph $G^{(t)} = (V^{(t)}, E^{(t)})$, where:
\begin{itemize}
    \item $V^{(t)} \subseteq V$ is the subset of agents active at time $t$.
    \item $E^{(t)} \subseteq V^{(t)} \times V^{(t)}$ is the set of directed edges at time $t$. A directed edge $(j \to i) \in E^{(t)}$ indicates that agent $i$ observes the output of agent $j$ at step $t$.
\end{itemize}
A \emph{graph update function} $\phi$ maps the previous graph, internal states, and inputs to the current graph, thereby evolving the topology over time:
\[
    G^{(t)} = \phi\left(G^{(t-1)}, \{(s_k, x_k)\}_{k \in V^{(t-1)}}\right).
\]

\end{definition}

Combining these agents and their evolving interactions, we now define the complete structure of a \emph{Multi-Agent AI System}.

\begin{definition}[Multi-Agent AI System]
\label{def:MAS}
An multi-agent AI system (MAS) is defined by a tuple:
\[
    \left(\mathcal{I}, \{(\mathcal{S}_i, \mathcal{X}_i, \mathcal{Y}_i, p_i)\}_{i \in \mathcal{I}}, \{(G^{(0)}, \phi)\}\right),
\]
where:
\begin{itemize}
    \item $\mathcal{I}$ is an index set of agents in the system.
    \item Each agent $i \in \mathcal{I}$ is defined by $(\mathcal{S}_i, \mathcal{X}_i, \mathcal{Y}_i, p_i)$ as per Definition~\ref{def:MAS_agent}.
    \item $G^{(0)}$ is the initial topology, and $\phi$ is the graph update function introduced in Definition~\ref{def:MAS_topology}.
\end{itemize}
\end{definition}

Based on the above definitions, and intuitively speaking, the MAS evolves over time by running the following sequence of operations at each time step~$t$:
\begin{enumerate}
    \item Each agent $i \in V^{(t-1)}$ receives input 
    \[
    x_i^{(t)} = \left(\left\{y_j^{(t-1)}\mid j \in V^{(t-1)}, (j \rightarrow i) \in E^{(t-1)}\right\}, \mathcal{F}_i^{(t)}\right).
    \]
    \item Each agent updates its internal state and generates an output according to its transition kernel $p_i$:
    \[
        (s_i^{(t)}, y_i^{(t)}) \sim p_i(\cdot \mid s_i^{(t-1)}, x_i^{(t)}).
    \]
    \item The topology evolves according to the graph update function $\phi$:
    \[
        G^{(t)} = \phi\left(G^{(t-1)}, \{(s_k, x_k)\}_{k \in V^{(t-1)}}\right).
    \]
\end{enumerate}

The iteration continues until a predefined stopping criterion is satisfied. The outputs produced by the MAS can be evaluated based on the specified evaluation metrics.

\begin{figure}[t]
\centering
\includegraphics[width=0.75\linewidth]{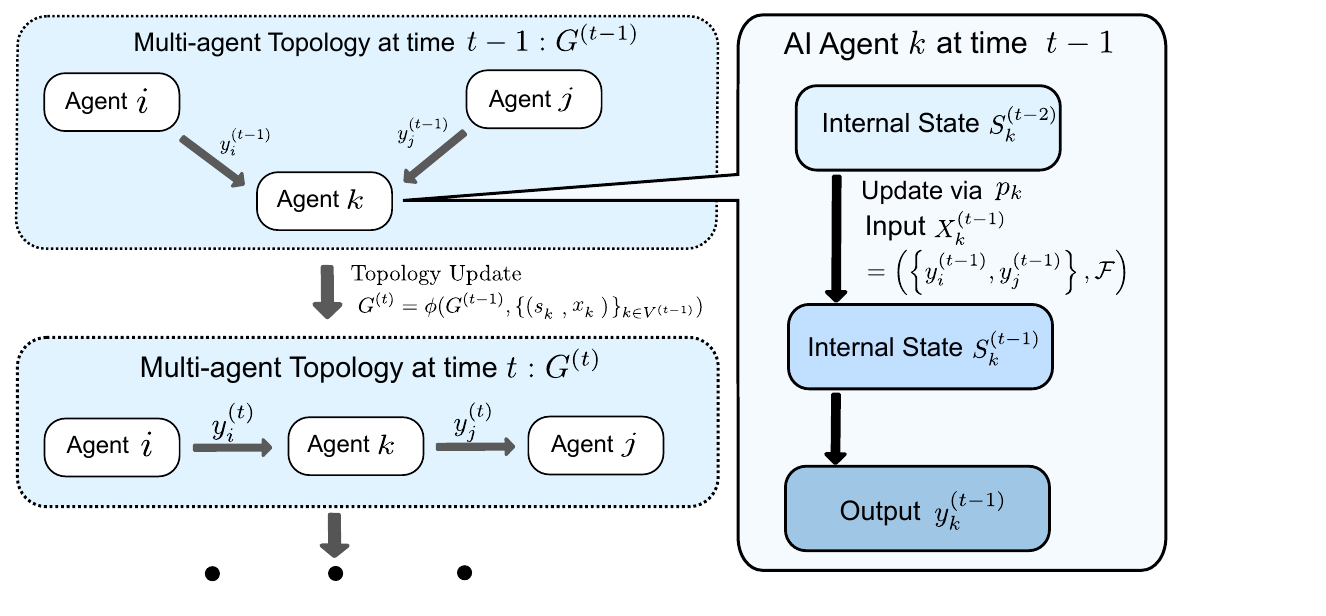}
\caption{
\textbf{Core update and topology dynamics in an MAS.} 
\textbf{Left:} Illustration of multi-agent topologies at time $t{-}1$ and $t$, showing how agent $k$ receives messages $y_i^{(t-1)}$ and $ y_j^{(t-1)}$ from connected agents and how the topology evolves via a graph update function $\phi$. 
\textbf{Right:} Update mechanism for agent $k$ at time $t{-}1$, where it updates its internal state based on received inputs and produces output $y_k^{(t-1)}$. 
}
\label{fig:MAS_core_elements}
\end{figure}

This proposed concise framework accommodates both symbolic and neural agents, and serves as the foundational structure for all subsequent modeling and experiments presented in this paper. Figure \ref{fig:MAS_core_elements} illustrates the core update and topology dynamics in an MAS.

Below, we exemplify the definitions of core elements in our framework with a concrete use case on flight booking.

\begin{cuteexample}[Definitions in Practice: A Flight Booking Use Case]
\label{ex:flightbooking}
Consider MAS designed for flight booking, consisting of the following agents:
\begin{itemize}
    \item \textbf{Agent 1 (User Interface)} has an internal state $\mathcal{S}_1$ (e.g., cached user queries), takes user-provided input $x_1^{(t)}$ containing origin, destination, travel dates, and preferences, and outputs structured query data $y_1^{(t)}$.

    \item \textbf{Agent 2 (Flight Search)} maintains an internal state $\mathcal{S}_2$ (e.g., cached flight availability), receives structured query input $x_2^{(t)}$ from Agent 1, and outputs specific flight options $y_2^{(t)}$, including airline, flight number, departure time, and price.

    \item \textbf{Agent 3 (Payment Processing)} holds internal state $\mathcal{S}_3$ (e.g., previous transaction data), receives input $x_3^{(t)}$ consisting of selected flight details from Agent 2 and payment information (exogenous feedback $\mathcal{F}_3^{(t)}$), and outputs payment confirmations or transaction failures $y_3^{(t)}$.

    \item \textbf{Agent 4 (Promotion)}, dynamically activated by the graph update function $\phi$, maintains an internal state $\mathcal{S}_4$ (e.g., user loyalty information), takes input $x_4^{(t)}$ consisting of confirmed booking details, determines promotion availability, and outputs advertisement $y_4^{(t)}$.
\end{itemize}

Each agent $i$ operates by updating its state $\mathcal{S}_i$ and generating output $y_i^{(t)}$ according to the probabilistic transition kernel $p_i$ based on inputs $x_i^{(t)}$ defined above. The MAS topology $G^{(t)}$ may form a directed chain, where Agent~1 forwards structured requests to Agent~2, and Agent~2 passes selected flight options to Agent~3 for payment processing. The graph update function $\phi$ dynamically adjusts this topology, for instance, activating Agent~4 to handle user-requested upgrades.
\end{cuteexample}

\noindent

In the next section, we show how \emph{feedback} mechanisms can further enrich this fundamental formulation by enabling agents to adjust their transition kernels $p_i$ or the topology $G^{(t)}$ in response to changing performance needs or adversarial threats.

\subsection{Feedback-Driven Updates}
\label{subsec_feedback_notation}

 Building on our formulation in Section ~\ref{subsec_mas_notation}, we highlight the central role of feedback in enabling the dynamic adaptability that distinguishes multi-agent AI systems from their single-agent counterparts. Rather than treating each agent’s function \( p_i \) as static, feedback can be incorporated to adjust the input-output relationship, effectively transforming \( p_i \) into a time-dependent function \( p_i^{(t)} \). 
 

\paragraph{How Feedback Updates Agents' Generation.}
At each time step $t$, an agent $i$ updates its state $s_i^{(t)}$ and output $y_i^{(t)}$ based on its input $x_i^{(t)}$, which consists of past agent outputs and exogenous feedback:
\[
p_i\left(s_i^{(t)}, y_i^{(t)}\mid s_i^{(t-1)}, x_i^{(t)}\right).
\]

Since $x_i^{(t)}$ includes exogenous feedback $\mathcal{F}_i^{(t)}$, receiving new feedback effectively conditions the transition kernel on additional information. This conditioning refines the agent’s belief about how to update its state and generate outputs, relative to when no feedback is incorporated. A natural way to represent this refinement is through Bayesian updating. Consider the transition kernel without
feedback as a prior:
\[
  p_i^{\text{prior}}
  \bigl(s_i^{(t)},y_i^{(t)}
        \mid s_i^{(t-1)},x_i^{(t)}\setminus \mathcal{F}_i^{(t)}\bigr),
\]
where
\(x_i^{(t)}\setminus \mathcal{F}_i^{(t)}
      = \{\,y_j^{(t-1)} \mid j\in V^{(t-1)},\;(j\rightarrow i)\in E^{(t-1)}\}\)
contains only the output from agents connected to agent $i$ at time $t-1$, and 
\(
x_i^{(t)}
=
\bigl(x_i^{(t)}\setminus \mathcal{F}_i^{(t)},\;
       \mathcal{F}_i^{(t)}\bigr).
\)
This transition kernel updates the state and output when there is no exogenous feedback. When the feedback \( \mathcal{F}_i^{(t)} \) arrives, it is incorporated through the Bayes rule:
\[
  p_i\!\left(s_i^{(t)},y_i^{(t)}
            \mid s_i^{(t-1)},x_i^{(t)}\right)
  \;\propto\;
  p\!\left(\mathcal{F}_i^{(t)}
           \mid y_i^{(t)},x_i^{(t)}\setminus \mathcal{F}_i^{(t)}\right)
  \cdot
  p_i^{\text{prior}}\!\left(s_i^{(t)},y_i^{(t)}
           \mid s_i^{(t-1)},x_i^{(t)}\setminus \mathcal{F}_i^{(t)}\right),
\]
where we assume that \(p(\mathcal{F}_i^{(t)}\mid y_i^{(t)},x_i^{(t)}\setminus \mathcal{F}_i^{(t)})= p(\mathcal{F}_i^{(t)}\mid s_i^{(t)},y_i^{(t)},s_i^{(t-1)},x_i^{(t)}\setminus \mathcal{F}_i^{(t)})\), i.e., the feedback depends only on the agent's output and input and not on its internal states. $p(\mathcal{F}_i^{(t)}\mid y_i^{(t)},x_i^{(t)}\setminus \mathcal{F}_i^{(t)})$ measures how
plausible the observed feedback is given output
\(y_i^{(t)}\) and input $x_i^{(t)}\setminus \mathcal{F}_i^{(t)}$. The resulting posterior \(p_i(\cdot \mid s_i^{(t-1)}, x_i^{(t)})\) exactly updates with feedback.

\paragraph{How Feedback Impacts Multi-Agent Topology.}
In a dynamic multi-agent AI system, feedback plays a crucial role in modifying not only individual agent functions but also the structure of inter-agent communication. When certain agents fail to function properly, produce low-quality outputs, or even generate adversarial responses, feedback mechanisms help redirect interactions to more reliable agents. This adaptability ensures that the system remains functional even in the presence of unreliable agents.  

As previously defined, the MAS topology evolves based on received feedback: 
\[
    G^{(t+1)} 
    \;=\; 
    \phi\Bigl(
        G^{(t)},\, 
        \{(s_k, y_k)\}_{k \in V^{(t)}},\,
       \{ \mathcal{F}_i^{(t)} \}_{i\in\mathcal{I}}
    \Bigr),
\]

where the function \( \phi \) determines how agent interactions evolve in response to feedback. If feedback indicates improved cooperation between certain agents, new edges may be added to \( G^{(t)} \), enhancing connectivity. Conversely, if feedback identifies unreliable agents, connections may be pruned, preventing their influence from degrading overall system performance. By dynamically adjusting the topology, feedback enables the MAS to maintain robustness and efficiency in evolving environments.

In the example below, we illustrate the feedback-driven updates in a multi-robot warehouse scenario.

\begin{cuteexample}[Practical Scenario: Multi-Robot Warehouse]
\label{ex:warehouse}
Consider a warehouse managed by a multi-agent AI system, where autonomous robots (agents) collaborate to pick and place items efficiently. Each robot $i$ updates its state $s_i^{(t)}$, processes input $x_i^{(t)}$, and generates output $y_i^{(t)}$ according to:
\[
p_i\left(s_i^{(t)}, y_i^{(t)}\mid s_i^{(t-1)}, x_i^{(t)}\right).
\]
Since $x_i^{(t)}$ includes exogenous feedback $\mathcal{F}_i^{(t)}$, the transition kernel $p_i$ is conditioned on additional information, refining the robot’s decisions and task coordination.

\begin{itemize}
    \item \textbf{Robot 1 (Inventory Scanner)} maintains an internal state $s_1^{(t)}$ tracking inventory data. It receives feedback $\mathcal{F}_1^{(t)}$ (e.g., new item arrivals or restock alerts). Its updated output $y_1^{(t)}$ provides an item availability list.

    \item \textbf{Robot 2 (Item Picker)} takes input $x_2^{(t)} = (y_1^{(t)}, \mathcal{F}_2^{(t)})$, where $y_1^{(t)}$ encodes inventory status and $\mathcal{F}_2^{(t)}$ includes human operator modifications or system reassignments. It updates its belief over available items and produces $y_2^{(t)}$ listing the items picked.

    \item \textbf{Robot 3 (Quality Checker)} receives $x_3^{(t)} = (y_2^{(t)}, \mathcal{F}_3^{(t)})$, where $y_2^{(t)}$ contains picked items and $\mathcal{F}_3^{(t)}$ consists of sensor-based quality assessments. It adjusts its quality classification, outputting $y_3^{(t)}$ as either "approve" or "reject."

    \item \textbf{Robot 4 (Order Dispatcher)} processes $x_4^{(t)} = (y_3^{(t)}, \mathcal{F}_4^{(t)})$, where $y_3^{(t)}$ provides inspected items and $\mathcal{F}_4^{(t)}$ contains external constraints (e.g., urgent orders). It dynamically updates shipment scheduling and outputs $y_4^{(t)}$ determining whether an order is shipped or held.
\end{itemize}

\paragraph{Impact on Topology.}
Initially, the MAS topology $G^{(t)}$ follows a directed sequence, where tasks flow from Robot 1 to Robot 4. However, upon receiving real-time feedback (e.g., damage alerts, shipment delays), the graph update function $\phi$ dynamically modifies agent connections. For instance:
\begin{itemize}
    \item If Robot 3 identifies defective items, $\phi$ reroutes them to Robot 2 for re-picking, altering $G^{(t)}$.
    \item If urgent orders are received ($\mathcal{F}_4^{(t)}$), Robot 4 may directly request prioritization from Robot 2, bypassing non-essential steps.
\end{itemize}
By iteratively conditioning the transition kernel on feedback and adjusting topology via $\phi$, the MAS adapts to operational changes, improving efficiency and coordination.
\end{cuteexample}

\begin{figure}[htbp]
\centering
\includegraphics[width=0.75\textwidth]{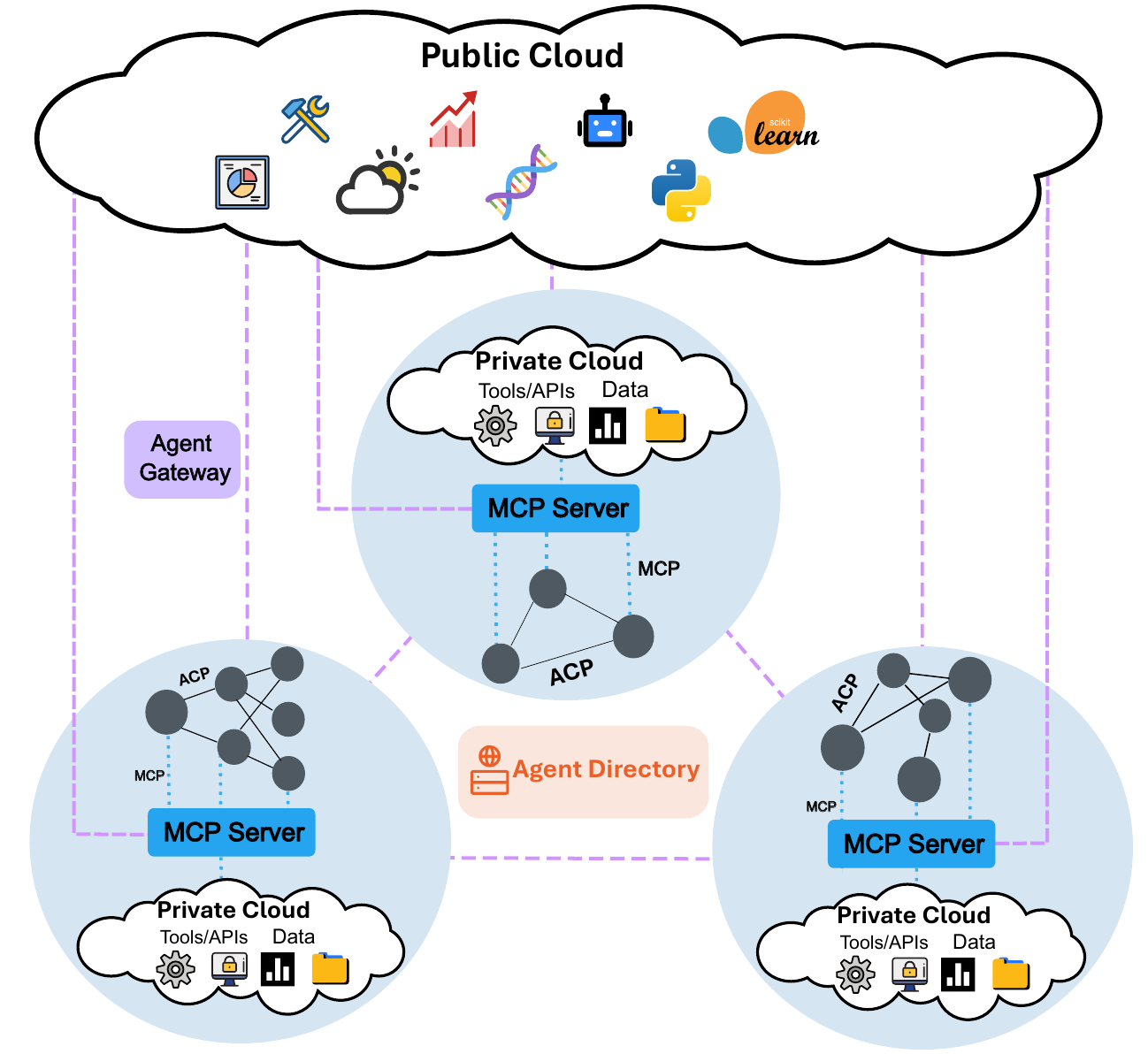}

\caption{
\textbf{Extension of MAS to Internet of MAS.}
A networked environment where agents can be registered, discovered, and composed much like modular web services with standardized interfaces and protocols. Specifically, agents register with an Agent Directory, analogous to the Domain Name System (DNS), and connect using the Agent Connect Protocol (ACP)~\cite{cisco_whitepaper_agntcy}, similar to service-level routing. Calls to external tools are structured via the Model Context Protocol (MCP)~\cite{AnthropicMCP2024} and executed through an MCP Server, like HTTP requests via an API gateway. All communication flows through the Agent Gateway, which functions as a secure proxy analogous to HTTPS gateways or service meshes. 
}

\label{fig:agntcy_lifecycle}
\end{figure}

\subsection{Extending MAS for Open and Networked Environments: Internet of MAS}

\label{AGNTCY}

The introduced MAS in Section~\ref{subsec_mas_notation} and Section~\ref{subsec_feedback_notation} describe how agents interact and evolve over time within a closed environment, such as a government agency, company, or research lab. However, in many real-world settings, agents are deployed across different systems and need to interact with others outside their original environment~\cite{Yan2025_MAS_comm,Gao2024_LLMAgentModeling, LangChain2022}. This raises new challenges such as how MAS can find each other, connect securely, and collaborate across system boundaries.

To address these challenges, we aim to extend the MAS framework to support open, networked environments where MAS from different systems can communicate and work together. This forms the basis of what we call the \emph{Internet of MAS}. To support this vision, the AGNTCY architecture~\cite{cisco_whitepaper_agntcy} was recently proposed, offering standardized protocols and components for MAS registration, discovery, coordination, and external tool access. Inspired by this design, we incorporate key mechanisms into our framework, as illustrated in Figure~\ref{fig:agntcy_lifecycle}.

To understand this extension, consider a simple analogy: think of each agent as a computer on the Internet, and each MAS as a local area network (LAN) or autonomous system (AS). For agents within and across MAS to interoperate, they must follow steps similar to how Internet-connected devices discover, communicate, and collaborate. First, agents must be made visible to other MAS. This is enabled by the \emph{Agent Directory}, which functions like the Domain Name System (DNS) or a search engine, allowing MAS agents to be registered, discovered, and queried across boundaries. Next, when agents attempt to connect, the \emph{Agent Connect Protocol (ACP)} defines their communication structure and network topology, much like DNS-based routing supports Internet-level service discovery. When MAS agents need to invoke external tools or services, this is handled via the \emph{Model Context Protocol (MCP)}, which defines standard message formats (analogous to HTTP or gRPC). These messages are routed through an \emph{MCP Server}, which selects and executes the correct tool. Finally, all inter-MAS communication flows through an \emph{Agent Gateway}, which serves as a secure proxy that enforces access policies, routes connections, and enables cloud-level function access. Each of these components supports MAS-level communication and collaboration over open, decentralized networks.

Below, we provide a mathematical formalization of the key components in the Internet of MAS. We first define Agent Directory, where agents can be registered, queried, and discovered based on their functional description.

\begin{definition}[Agent Directory]
Let $\mathcal{I}$ be a set of agent identifiers. An \emph{Agent Directory} is a map
\[
\mathcal{D} : \mathcal{I} \to \mathcal{M},
\]
where each $\mathcal{D}(i) \in \mathcal{M}$ contains structured metadata describing the input/output interface, declared capabilities, and version of agent $i$.
\end{definition}

Once agents are available on the Agent Directory, ACP determines which agents can communicate with others based on whether the output message formats of one agent are accepted as input by another.

\begin{definition}[Agent Connect Protocol (ACP)]
Let $V \subseteq \mathcal{I}$ be a set of agents, where each agent $i \in V$ declares:
\begin{itemize}
    \item an output type set $\mathcal{Y}_i$ (the types of messages it can emit),
    \item an input type set $\mathcal{X}_i$ (the types of messages it can consume).
\end{itemize}
The \emph{Agent Connect Protocol (ACP)} defines a directed graph $G = (V, E)$, where
\[
(j \to i) \in E \quad \text{iff} \quad \mathcal{Y}_j \cap \mathcal{X}_i \ne \emptyset.
\]
\end{definition}

While ACP governs agent-to-agent communication, MCP addresses how agents interact with external tools and services. Each MCP message includes a header that specifies which tool to call, and a body that provides the required parameters. The MCP Server reads the header to select the correct tool and passes the body to it as input.

\begin{definition}[Model Context Protocol and Server]
Let $\mathcal{M} = \mathcal{H} \times \mathcal{B}$ be the MCP message space, where:
\begin{itemize}
    \item $\mathcal{H}$ is the space of headers (e.g., task type, auth token),
    \item $\mathcal{B}$ is the space of message bodies (e.g., structured parameters).
\end{itemize}

An \emph{MCP Server} is a function:
\[
\mathcal{S}_{\mathrm{mcp}} : \mathcal{M} \to \mathcal{O}, \quad \text{defined as } \mathcal{S}_{\mathrm{mcp}}(h, b) := f_h(b),
\]
where $f_h \in \mathcal{F}$ is the function specified by header $h$.
\end{definition}

Finally, the Agent Gateway serves as the central checkpoint for inter-agent communication across systems. It enforces security and access control by determining whether messages between agents are allowed to pass, based on predefined system policies.

\begin{definition}[Agent Gateway]
An \emph{Agent Gateway} controls whether a message between two agents is allowed to pass. It uses a simple rule:
\[
\Pi(\text{src}, \text{dst}, m) =
\begin{cases}
1 & \text{allowed, if the message is permitted by system policies} \\
0 & \text{otherwise}
\end{cases}
\]
If $\Pi$ returns 1, the gateway applies a processing function
\[
\mathcal{G}(\text{src}, \text{dst}, m) = m',
\]
where:
\begin{itemize}
    \item $\text{src} \in \mathcal{I}$ is the sending agent,
    \item $\text{dst} \in \mathcal{I}$ is the receiving agent,
    \item $m \in \mathcal{M}$ is the original message (e.g., an MCP message),
    \item $m' \in \mathcal{M}$ is the processed message after optional formatting, filtering, or routing.
\end{itemize}

If $\Pi$ returns 0, the message is blocked and not delivered.
\end{definition}

\section{Is Multi-Agent More Effective Than Single-Agent?}
\label{sec:effectiveness}

The adoption of MAS has been widely explored in various fields, with proposed benefits including improved computational efficiency~\cite{Panait2005}, enhanced adaptability~\cite{Tan1993}, and greater robustness~\cite{Vinyals2019}. 

However, whether MAS is fundamentally more effective than single-agent systems remains an open question.  Regarding the underlying mechanisms that may contribute to MAS advantages, different perspectives exist. Some studies argue that decomposing tasks among multiple agents enables parallel processing, thereby reducing computational complexity~\cite{Foerster2016Comm}. Others suggest that distributed decision-making improves robustness against agent failures ~\cite{HernandezLeal2019}, while additional work highlights the role of self-organization in dynamic task allocation~\cite{Leibo2017}. 
Despite these theoretical advantages, MAS does not universally outperform single-agent approaches. Coordination overhead~\cite{Stone2000}, conflicting agent objectives~\cite{Shoham2009}, and suboptimal communication structures~\cite{Lowe2017} can introduce inefficiencies. Identifying the fundamental principles that govern when MAS is effective and when it fails is critical for designing robust and scalable multi-agent architectures.

\begin{figure}[htbp]
\centering
\includegraphics[width=0.75\textwidth]{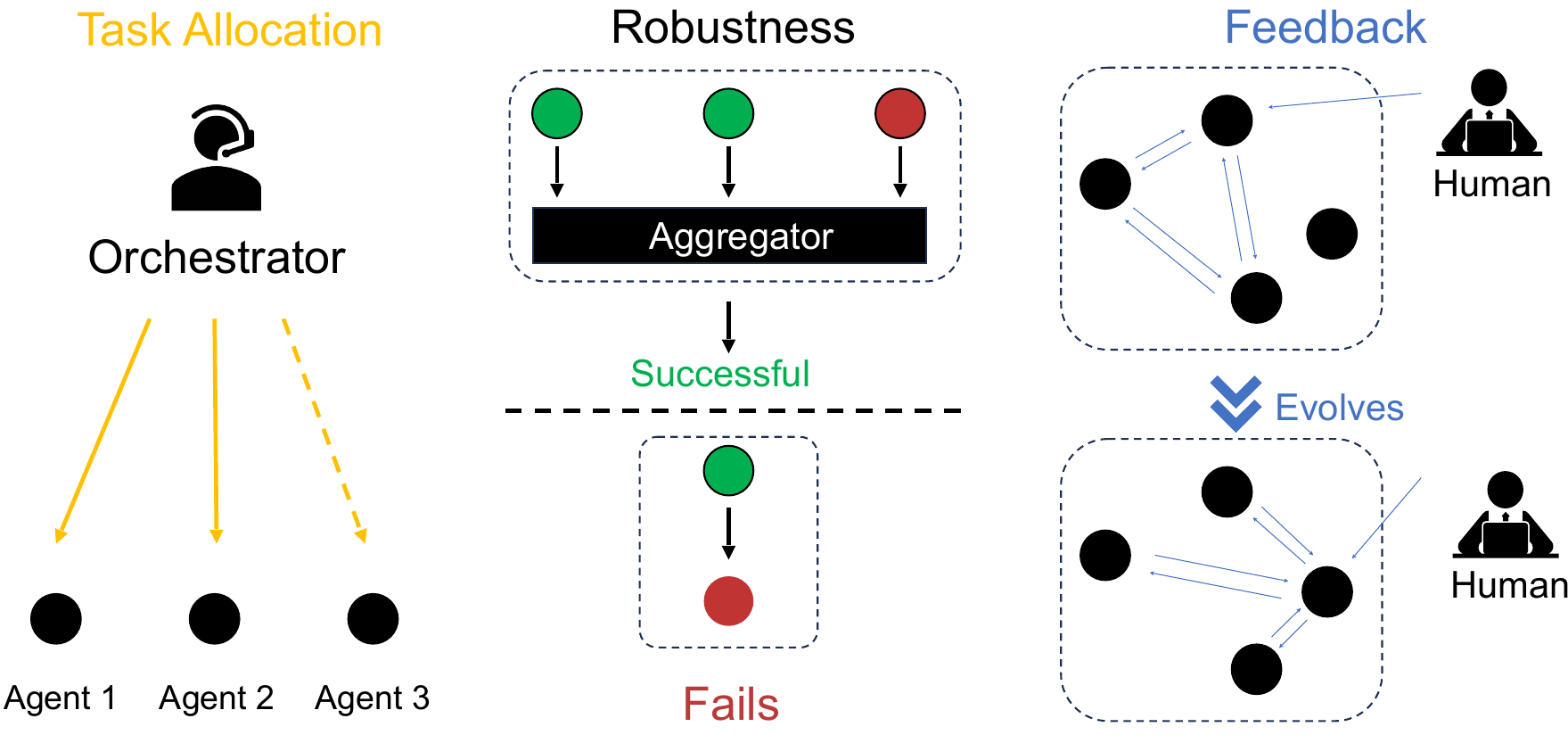}
\caption{\textbf{Overview of our position on effectiveness of MAS.} Nodes represent agents. \textbf{Left}: MAS can dynamically allocate tasks across agents to adapt to varying requirements. \textbf{Middle}: MAS may or may not exhibit robustness. In the upper example, even if some agents fail, the system can still succeed through a vote aggregator. In contrast, the lower example shows a case where the failure of any single agent causes the entire system to fail. \textbf{Right}: MAS benefits from both external feedback (e.g., from humans) and internal feedback via inter-agent communication. Feedback can even drive the system to evolve its topology. }
\label{fig:effectiveness}
\end{figure}

This motivates the research question:    \emph{Under what conditions does a multi-agent AI system outperform a single-agent one, and what are the principles that govern its success or failure?}
To  evaluate this question, we frame our analysis around three progressively connected perspectives: task allocation, robustness, and feedback integration. Task allocation considers how MAS can dynamically divide and coordinate tasks more effectively than single agents, drawing from divide-and-conquer strategies. Robustness examines whether these benefits hold under uncertainty and failure, relating closely to ensemble learning principles. Feedback integration explores how MAS adapt over time through internal and external signals, leveraging Bayesian approaches. These perspectives form a layered framework for assessing when and why MAS can outperform single-agent systems. See Figure \ref{fig:effectiveness} for an illustration of the three perspectives.

\subsection{Perspective One: The Potential Benefit of Multi-Agent AI Systems in Task Allocation}

%

Our analysis of task allocation is a reminiscence of the classical “divide-and-conquer” paradigm. This paradigm addresses complex computational problems by decomposing them into smaller subproblems, solving each independently, and then combining the results~\cite{Aho1974}. Traditionally, subtasks are determined in advance and assigned to computational agents in a static manner~\cite{Stone2000, Shoham2009}. While this strategy works effectively in stable and predictable environments, it often struggles under dynamic conditions such as workload fluctuations, agent performance variations, or external disruptions~\cite{Leibo2017}.

Multi-agent AI systems (MAS) address the limitations in “divide-and-conquer” by allowing agents to dynamically adjust task allocations based on real-time feedback and environmental changes~\cite{Vinyals2019}. Instead of relying on fixed task partitions, MAS adaptively refine the distribution of workloads among agents in response to current conditions. This capability naturally leads to a central question: 
\emph{Can MAS autonomously identify optimal task distributions in real-time without predefined problem decompositions?}  

Answering this question shifts the focus from static problem divisions to dynamic, iterative decision-making. Agents continuously revise their task assignments using ongoing feedback on their own performance and environmental state~\cite{Foerster2016,Lowe2017}. Thus, the main challenge for MAS becomes effectively implementing dynamic task reallocation strategies that balance responsiveness, flexibility, and computational efficiency~\cite{Vinyals2019}.

\vspace{0.1cm}
\textbf{A New Opportunity: Self-Organized Task Allocation in MAS Beyond Human-Specified Decompositions.}
In traditional systems, humans have to decide in advance how to divide a complex task into smaller subtasks and assign them to different agents. This approach works well in stable environments. But in dynamic settings, where workloads shift, agents perform differently over time, or external feedback changes, fixed task assignment often becomes inefficient or even fails.

In this work, we propose a new perspective: \emph{can MAS learn how to divide and assign tasks on its own, without relying on any human-specified decomposition?} This capability, if realized, would allow MAS to handle environments that are too complex or unpredictable for human pre-programming. 

We describe the structure of an MAS as a dynamic interaction graph \( G^{(t)} = (V^{(t)}, E^{(t)}) \), where nodes are agents and edges describe communication links. Each agent receives input from its neighbors and the environment, updates its internal state, and outputs a response:
\[
    (s_i^{(t)}, y_i^{(t)}) \sim p_i(s_i^{(t)}, y_i^{(t)} \mid s_i^{(t-1)}, x_i^{(t)}).
\]

Unlike static systems, the interaction structure \( G^{(t)} \) can change over time. The system uses a graph update rule:
\[
    G^{(t)} = \phi(G^{(t-1)}, \{(s_k^{(t-1)}, x_k^{(t-1)})\}_{k \in V^{(t-1)}}),
\]
so that agents can rewire their communication links based on performance, feedback, or needs.

The dynamic adaptability allows MAS to go beyond fixed task splits. Instead of following a hard-coded divide-and-conquer scheme, agents \emph{self-organize} based on real-time feedback. The system iteratively explores and reshapes its internal structure to better match task demands --- something that would be \emph{infeasible through human pre-specification or static programming}.

\vspace{0.1cm}
\textbf{An Underexplored Challenge: Error Amplification in Self-Organizing Task Allocation.}
The ability of MAS to adaptively organize and rewire their task assignments could be powerful, but it also comes with risks. In traditional systems with fixed task assignment, errors tend to stay local. However, the dynamic structure of MAS means that a single mistake can quickly spread through the entire system. To illustrate this, consider the previous setting where agents continuously update their communication links based on performance and feedback. While this helps the system adapt to new conditions, it can also lead to fragile coordination if not handled carefully. For example, suppose agent \( h \) becomes a central connector, or “hub”, that many other agents depend on for information. If \( h \) makes a mistake, its output may mislead every agent connected to it. Formally, the input to a dependent agent \( j \) can be written as:
\[
x_j^{(t)} = \{ y_h^{(t-1)} \} \cup \{ y_k^{(t-1)} \mid k \in \mathcal{N}_j^{(t-1)} \setminus \{h\} \} \cup \mathcal{F}_j^{(t)}.
\]
Even if only \( h \) fails, its error may be passed down to many others in the next step. Over time, these small mistakes can snowball into system-wide failures.


This raises a critical challenge: how can we design MAS that not only adapt quickly but also remain robust to individual agent failures? Unlike static architectures, where failures remain localized, in MAS, poor connectivity design can amplify errors system-wide. Thus, a future research direction is to develop mechanisms that optimize \( \phi \) not only for efficiency but also for stability, ensuring that the benefits of adaptive task allocation do not come at the cost of increased fragility.

\subsection{Perspective Two: The Limits of Multi-Agent AI Systems in Robustness}
\label{sec:robustness_redundancy}

Beyond the benefit from dynamic topology design in MAS, another widely held belief is that MAS are inherently more reliable than single-agent systems. A common intuition is that redundancy improves reliability: if one agent fails, others can take over its role. By having multiple agents capable of handling similar tasks, the system is expected to tolerate isolated errors more effectively. This idea draws inspiration from ensemble learning in machine learning \cite{Dietterich2000, Polikar2006}, where multiple models (often termed ``weak learners'') are combined to improve overall decision accuracy. Classical ensemble methods such as Bagging \cite{Breiman1996}, Boosting \cite{Freund1997}, and Random Forests \cite{Breiman2001} leverage diverse predictions to mitigate individual errors, theoretically leading to more stable performance. 

\paragraph{A Common Belief: Redundancy Makes MAS Robust by Majority Voting.}
To analyze the robustness from redundancy, consider an MAS composed of \( k \) agents, where each agent \( g_i \) independently makes a binary decision \( y_i \in \{0,1\} \). Suppose each agent correctly predicts the true label with probability \( 1 - e_i \), where \( e_i \) represents the error rate. The system aggregates individual decisions via a majority vote function:

\[
F(y_1, y_2, \dots, y_k) =
\begin{cases}
  1, & \text{if } \sum_{i=1}^{k} X_i \geq \frac{k}{2},\\
  0, & \text{otherwise},
\end{cases}
\]

where \( X_i \sim \text{Bernoulli}(1 - e_i) \) indicates whether agent \( g_i \) predicts correctly.

By Hoeffding’s inequality \cite{Hoeffding1963}, the probability of an erroneous final decision is bounded as:

\[
\mathbb{P}\left( \frac{1}{k} \sum_{i=1}^{k} X_i < \frac{1}{2} \right) 
\leq \exp\left\{ -2k \left( (1 - \bar{e}) - \frac{1}{2} \right)^2 \right\}
\]

where \( \bar{e} = \frac{1}{k} \sum_{i=1}^{k} e_i \) is the average agent error rate. This suggests that as \( k \) increases, the system’s robustness improves exponentially, but only under the assumption that agent errors are independent.

\vspace{0.1cm}
\textbf{The Fundamental Problem: Assumptions Underlying the Robustness Could Fail.}
The theoretical result above suggests that adding more agents improves robustness. However, this relies on the assumption that the training data behind agents have a certain level of independence and the training objective is somewhat aligned.  
This leads to a central question: \emph{When does redundancy actually help, and when does it fail?} In the following, we highlight two common failure modes, high dependency among agents and misalignment of goals,  that can undermine the expected benefits of redundancy. 

\begin{tcolorbox}[
  colback=blue!10!white,
  colframe=blue!80!black,
  boxrule=0.5pt,
  arc=1mm,
  boxsep=1pt,
  left=1mm, right=1mm, top=0.5mm, bottom=0.5mm,
  enhanced,
  sharp corners,
  on line,
  box align=base,
  show bounding box
]
\textbf{Failure Mode 1: High Dependency Among Agents.}
\end{tcolorbox}

The majority vote strategy assumes that agents make independent errors. But in practice, agents may share training data, interact closely, or operate in similar environments. These factors can lead to highly correlated decisions.

Suppose agents \( g_i \) and \( g_j \) make decisions with correlation \( \lambda \). Then their agreement is:
\[
\mathbb{E}[X_i X_j] = (1 - e_i)(1 - e_j) + \lambda \sqrt{(1 - e_i)e_i} \cdot \sqrt{(1 - e_j)e_j}
\]

When \( \lambda \to 1 \), their predictions become nearly identical, and the group behaves like a single agent. In this case, increasing the number of agents does not improve robustness. Instead, it may just reinforce the same bias. This problem is common in large-scale MAS trained on similar data sources.

\begin{tcolorbox}[
  colback=blue!10!white,
  colframe=blue!80!black,
  boxrule=0.5pt,
  arc=1mm,
  boxsep=1pt,
  left=1mm, right=1mm, top=0.5mm, bottom=0.5mm,
  enhanced,
  sharp corners,
  on line,
  box align=base,
  show bounding box
]
\textbf{Failure Mode 2: Misalignment of Goals.}
\end{tcolorbox}

Even if agent failures are independent, robustness may still fail if a subset of agents actively work against the system's objective. Consider a MAS where agents are divided into two groups:

\[
\underbrace{\{g_1, \dots, g_{k_0}\}}_{\text{Aligned Group (correct w.p. } 1 - e_0)}
\cup
\underbrace{\{g_{k_0+1}, \dots, g_{k}\}}_{\text{Misaligned Group (correct w.p. } 1 - e_1)}.
\]

If \( k \gg k_0 \) or \( e_1 \gg e_0 \), then the misaligned group dominates the final decision, nullifying the benefits of redundancy. Specifically, the probability of an incorrect majority vote follows:

\[
\mathbb{P}\left( \sum_{i=1}^{k_0} X_i + \sum_{i=k_0+1}^{k} X_i < \frac{k}{2} \right),
\]

which increases significantly as \( k_1 \) grows larger or as misaligned agents deliberately introduce errors. Unlike independent errors, misalignment leads to systematic failure, which redundancy cannot mitigate.

\vspace{0.1cm}
\textbf{The Challenges in Real-World: Overlapping Training Data.}
In practice, the above two key failure modes: high dependency and misalignment of goals, often stem from a common root cause: overlapping training data. Unlike traditional ensemble learning, which promotes diversity through data resampling, MAS agents are frequently trained on large-scale datasets with substantial content similarity. For instance, recent studies \cite{Gao2020Pile,RedPajama2023} indicate that LLMs often share significant portions of training data, commonly drawn from open-source corpora such as The Pile, Common Crawl, and StarCoder \cite{HuggingFace2024}. This overlap can induce systematic correlations in agents' outputs. In the extreme case where all agents are trained on the same data, MAS effectively collapses to a single-agent system.
These observations challenge the prevailing belief that the diversity of LLMs inherently enhances robustness in MAS. 


\subsubsection{Empirical Study: Redundancy by Ensemble Voting on Overlapping Training Data}

To evaluate whether redundancy in MAS genuinely enhances robustness, we conduct a controlled experiment using ensemble-based majority voting over independently trained agents. Specifically, we use the \texttt{Covtype} dataset from \texttt{scikit-learn}, a commonly used classification dataset. Each agent is trained on a small subset of data, and we vary two key parameters: \textbf{Overlap Ratio} $\rho \in [0, 1]$: The proportion of each agent's training data that overlaps with others. Higher $\rho$ implies stronger similarity between agents due to common data sources. \textbf{Number of Agents} $k$: The ensemble size, i.e., the number of individually trained agents that will participate in majority voting to construct the final decision. We simulate $k$ agents, each trained on $100$ samples (with shared and unique portions controlled by $\rho$), and evaluate the ensemble performance via majority voting with random tie-breaking.

\begin{figure}[t]
\centering
\includegraphics[width=1.01\linewidth]{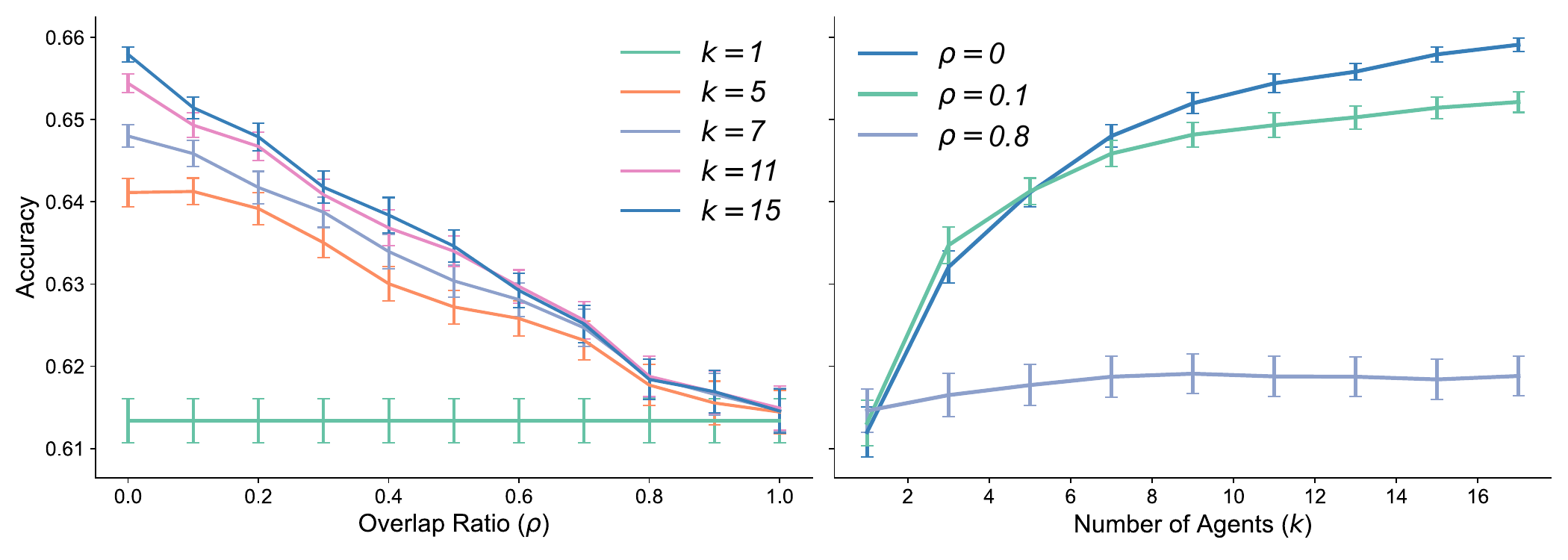}
\caption{
\textbf{Ensemble accuracy under varying overlap and ensemble size.}
\textbf{Left:} Accuracy decreases with overlap ratio $\rho$ for $k > 1$, showing that high overlap reduces diversity and harms ensemble performance. When $\rho = 1$, all agents behave similarly, matching the $k = 1$ baseline.
\textbf{Right:} When $\rho = 0$, accuracy improves clearly with larger $k$, illustrating the benefit of redundancy under independent errors. With increasing $\rho$, this improvement diminishes due to reduced diversity.
}

\label{fig:ensemble_experiment}
\end{figure}

Figure~\ref{fig:ensemble_experiment} summarizes our experimental results. The left panel plots ensemble accuracy against the overlap ratio \(\rho\), under varying ensemble sizes \(k \in \{1, 5, 7, 11, 15\}\). We observe a clear trend: for all \(k > 1\), ensemble accuracy consistently declines as \(\rho\) increases. This demonstrates that excessive data overlap across agents undermines the benefits of ensembling. In particular, when \(\rho = 1\), each agent is trained on exactly the same data, and the ensemble behaves like a single model repeated \(k\) times. As in our theoretical discussion, the ensemble in this setting fails to gain robustness, since agents make highly correlated decisions. This is an empirical instance of the failure mode where \(\rho \to 1\) leads to collapsed diversity. The \(k=1\) curve serves as a baseline that reflects the performance of a single-agent system. This comparison illustrates that increased redundancy does not translate to improved robustness when diversity among agents is compromised by overlapping training data.

The right panel of Figure~\ref{fig:ensemble_experiment} plots ensemble accuracy against the number of agents \(k\), under three different overlap ratios \(\rho \in \{0, 0.1, 0.8\}\). When \(\rho = 0\), agents are trained on fully disjoint datasets, and the resulting predictions are approximately independent. In this setting, we observe a clear and steady increase in ensemble accuracy as \(k\) grows, accompanied by a noticeable reduction in the variance (error bars) of the prediction accuracy. This behavior reflects improved robustness due to aggregation of independent decisions and matches the guarantee provided by Hoeffding’s inequality
. Hence, under minimal data overlap, redundancy in MAS exhibits its strongest benefits: increasing both accuracy and stability of the system. In comparison, when \(\rho = 0.1\) or \(\rho = 0.8\), agents share a non-negligible portion of their training data, and their predictions become partially correlated. As a result, especially when \(\rho = 0.8\), the accuracy improvements with increasing \(k\) are reduced or even saturated. While overlap ratio is not a direct measure of prediction correlation, it serves as a practical proxy. These results highlight that the robustness gains from redundancy are most effective when agent decisions are close to independent, which in practice requires careful control over training data overlap.

\vspace{0.5em}

These results show that adding more agents only improves robustness when their predictions are not too similar. In real-world MAS, if agents are trained on highly overlapping data, they tend to make similar mistakes, and voting does not help. To truly benefit from redundancy, it is important to ensure enough diversity in how agents are trained.

\subsection{Perspective Three: Challenge of Incorporating Feedback in Multi-Agent AI Systems}\label{sec:feedback}

MAS offer a special advantage over single-agent models by leveraging structured feedback loops to refine decision-making dynamically. Unlike single-agent systems, where feedback primarily influences a single trajectory, MAS enables agents to condition their updates based on both endogenous and exogenous feedback, leading to improved adaptability and collaboration. This conditioning mechanism allows for continuous refinement of agent decisions, making MAS inherently more flexible than single-agent systems.


Feedback not only refines agent behavior but also reshapes the MAS topology. If feedback signals indicate improved performance, edges can be reinforced, improving cooperation among agents. Conversely, poor feedback can prompt topology pruning, preventing unreliable agents from negatively influencing the system. This adaptability ensures robust coordination in MAS, setting it apart from static single-agent systems.

\vspace{0.1cm}
\textbf{The Fundamental Challenge: Effective Feedback Integration.}
A central question in MAS is how to effectively integrate feedback. To analyze this, we classify feedback into the following two types: endogenous feedback and exogenous feedback. Endogenous feedback is internal agent-to-agent signals, such as an agent detecting a failure in its task (e.g., a code-writing agent producing an error that affects subsequent agents). Exogenous feedback is human interventions or external environmental signals, which remain largely underexplored in MAS but have been studied extensively in reinforcement learning with human feedback (RLHF)~\citep{griffith2013policy,arumugam2019deep,bai2022training,dai2024safe}.

For endogenous feedback, the simplest approach for integration is direct propagation: the agent appends its output $y_i^{(t)}$ to the next agent's input $x_{i+1}^{(t+1)}$, allowing downstream agents to adjust accordingly. This mechanism aligns well with the Bayesian conditioning perspective discussed earlier.

For exogenous feedback, integration is less straightforward. MAS feedback mechanisms can be viewed as generalizing RLHF in MAS with our formulation. Let 
$\theta$ denote the policy parameters, and 
$D$ denote the observed human feedback data, RLHF typically updates policies using human preference data by applying an update rule of the form:
\[
p(\theta | D) \propto p(D | \theta) \cdot p(\theta),
\]
where the likelihood function $p(D | \theta)$ is often modeled as an exponential function of a reward model:
\[
p(D | \theta) \propto e^{\sum_{i} \lambda_i r_i(\theta)}.
\]
This form mirrors the Bayesian conditioning update in MAS, where $e^{-\text{loss}}$ in RLHF plays a role similar to the likelihood function $p(\mathcal{F}_i^{(t)}\mid y_i^{(t)},x_i^{(t)}\setminus \mathcal{F}_i^{(t)})$ in MAS feedback updates. While existing RLHF methods focus on single-agent learning, our formulation suggests that MAS feedback mechanisms can be viewed as generalizing these approaches.

Despite its potential, the effective application of RLHF in MAS remains underexplored. Existing literature on multi-agent reinforcement learning highlights structured inter-agent feedback mechanisms \cite{Zhang2019MARLOverview}, but integrating human feedback in MAS settings requires novel approaches. Notably, the challenge of real-time feedback propagation across multiple agents introduces additional complexities compared to standard RLHF frameworks.

\vspace{0.1cm}
\textbf{The Pitfall of Poor Feedback Integration: Reducing to a Single-Agent System.}
If feedback is not effectively injected into MAS, the system risks degenerating into a single-agent paradigm.
Formally, consider an MAS where agent interactions follow a directed graph $G^{(t)} = (V^{(t)}, E^{(t)})$. If feedback updates do not sufficiently alter agent transitions, then for all agents $i$, $\phi(G^{(t+1)}) \approx G^{(t)}$, which effectively collapses the system into a static single-agent model. Thus, ensuring that feedback dynamically influences $\phi$ is crucial for preserving the advantages of MAS.

To fully leverage feedback in MAS, future research should address the following aspects. First, it is essential to develop advanced techniques for aggregating multi-source, multi-modality feedback, dynamically reconciling both exogenous human input and endogenous agent signals. Second, the extension of RLHF frameworks from single-agent to multi-agent contexts will be critical for ensuring that MAS align with multi-dimensional human values or preferences~\cite{wang2025map}. Finally, future MAS architectures could be designed to enable interpretable agent-to-agent feedback propagation.

\begin{cuteexample}[Effectiveness Comparison of Two-Agent and Single-Agent Systems: A Data Science Case]
\label{ex:two_vs_single_agent_ds}
Consider a collaborative data science assistant implemented in two different ways:
\begin{itemize}
    \item \textbf{Single-Agent System}: A single agent attempts to design and execute the entire data science pipeline, including preprocessing, modeling, and evaluation.
    \item \textbf{Two-Agent System}: The task is decomposed into two stages, handled by distinct agents that exchange feedback.
\end{itemize}

\textbf{Two-Agent Decomposition:}
\begin{itemize}
    \item \textbf{Agent 1 (Workflow Proposal)}: Proposes a data science workflow \( y_1 \) based on task description and data metadata (such as suggesting imputation + random forest + cross-validation).
    \item \textbf{Agent 2 (Execution and Refinement)}: Executes the proposed pipeline, evaluates performance, and refines the workflow \( y_2 \) by tuning hyperparameters or modifying components.
\end{itemize}

Each agent \( i \) maintains an internal state \( s_i \), receives input \( x_i^{(t)} \), and outputs \( y_i^{(t)} \). The MAS topology \( G^{(t)} \) is sequential: \textbf{Agent~1} proposes a workflow passed to \textbf{Agent~2} for refinement and finalization.

\textbf{Cost Comparison:}
\begin{itemize}
    \item \textbf{Single-Agent Cost:} The agent must explore a large joint space of pipelines and parameters:
    \[
    \text{Cost}_{\text{single}} = c_{\text{single}} \cdot \mathbb{E}[\text{Steps}_{\text{multi-dim}}],
    \]
    where \( c_{\text{single}} \) is the cost per step (e.g., running a full pipeline), and \( \text{Steps}_{\text{multi-dim}} \) is the expected number of steps needed for end-to-end optimization.
    
    \item \textbf{Two-Agent Cost:} Task decomposition reduces the search space:
    \[
    \text{Cost}_{\text{two-agent}} = c_1 \cdot \mathbb{E}[\text{Steps}_{y_1}] + (c_1 \cdot \mathbb{E}[\text{Steps}_{y_1}] + c_2) \cdot \mathbb{E}[\text{Steps}_{y_2 \mid y_1}] + \delta_{\text{comm}},
    \]
    where \( c_1 \) and \( c_2 \) are the per-step costs for proposing and executing pipelines respectively, and \( \delta_{\text{comm}} \) captures the overhead of exchanging and interpreting workflow proposals.
\end{itemize}

\textbf{Conditions for Cost Reduction:}
The two-agent system is advantageous when:
\begin{itemize}
    \item \textbf{Step Reduction:}
    \[
    \mathbb{E}[\text{Steps}_{y_1}] + \mathbb{E}[\text{Steps}_{y_2 \mid y_1}] \ll \mathbb{E}[\text{Steps}_{\text{multi-dim}}].
    \]
    \item \textbf{Low Communication Overhead:}
    \[
    \delta_{\text{comm}} \ll c_2 \cdot \mathbb{E}[\text{Steps}_{y_2 \mid y_1}].
    \]
\end{itemize}

\textbf{Conclusion:}
The two-agent strategy improves efficiency when decomposition reduces the pipeline search dimensionality, and the communication cost between agents remains small. The cost advantage is determined by:
\[
\Delta \text{Cost} = \text{Cost}_{\text{single}} - \text{Cost}_{\text{two-agent}} > 0.
\]
This holds when:
\[
\frac{\mathbb{E}[\text{Steps}_{\text{multi-dim}}]}{\mathbb{E}[\text{Steps}_{y_1}] + \mathbb{E}[\text{Steps}_{y_2 \mid y_1}]} \gg \frac{c_1 + c_2 + \delta_{\text{comm}} / \mathbb{E}[\text{Steps}_{y_2 \mid y_1}]}{c_{\text{single}}}.
\]
This example illustrates how dividing a data science task into proposal and refinement stages enables more efficient search and decision-making under constrained computational budgets.
\end{cuteexample}

\section{Safety in Multi-Agent AI Systems}
\label{sec:mas_safety}

As AI systems become more complex and widely deployed, ensuring their safety has emerged as a critical challenge. In single-agent settings, safety issues such as adversarial attacks~\cite{szegedy2013intriguing,goodfellow2014explaining, yuan2019adversarial}, jailbreaking attacks~\cite{wei2023jailbroken,Liu2023PromptInjection}, data or model privacy~\cite{evfimievski2003limiting, dwork2006calibrating, ding2021interval, wang2024modelprivacy}, and especially backdoor attacks~\cite{xian2023unified, xian2023adaptability, wang2024demystifying} have been extensively studied and partially mitigated. As these foundational safety efforts mature, an essential extension involves moving beyond single-agent 
settings to scenarios where multiple agents operate within a shared environment. This shift inevitably heightens complexity, as 
interactions between agents may introduce new failure modes or exacerbate existing vulnerabilities.

Unlike single-agent architectures, where failures tend to remain localized, 
MAS exhibit complex interdependencies that can 
propagate or even amplify existing vulnerabilities 
\cite{dafoe2020open}. These interdependencies significantly complicate efforts to ensure robustness, reliability, and security across the entire system. This raises a critical question:
\begin{quote}
  \emph{Do multi-agent AI systems amplify or mitigate safety risks compared to single-agent systems, 
  and under what circumstances?}
\end{quote}

Next, we introduce a framework to study these questions, present toy experiments, and use backdoor attacks to illustrate how different MAS can impact amplification or attenuation of vulnerabilities.

\subsection{Formalizing Vulnerability Propagation in MAS}

\textbf{General Notion of Vulnerability.}
A single-agent system \( F: \mathcal{X} \to \mathcal{Y} \) is said to exhibit a \emph{vulnerability} if there is a small perturbation \( \delta \), either in the input or model configuration, such that
\[
\mathcal{V}(F; \mathbb{P}, \delta) := \mathcal{M}(F; \mathbb{P}) - \mathcal{M}((F; \mathbb{P})_{\delta})
\]
is substantially large. Here, \( \mathcal{M}(F; \mathbb{P}) = \mathbb{E}_{X \sim \mathbb{P}} [M(F(X), X)] \) is a task-specific performance metric, the larger the better. The perturbation \( \delta \) may reflect changes in the input space (e.g., adversarial examples, trigger injections) or model structure (e.g., poisoned weights, latent neuron triggers). The notation \( \mathcal{M}((F; \mathbb{P})_{\delta}) \) denotes the performance of the system after perturbation \( \delta \) is applied.

\vspace{0.1cm}
\textbf{Extension to Multi-Agent Pipelines.}
In MAS, multiple agents \( \{g_1, g_2, \dots, g_n\} \) may process data sequentially or in parallel, where each \( g_i: \mathcal{X}_i \to \mathcal{X}_{i+1} \). The full system forms a pipeline:
$
F_{\text{MAS}} = g_n \circ \dots \circ g_1.
$
The propagation of vulnerabilities across such a pipeline becomes a key concern: earlier agents may amplify or mitigate vulnerabilities based on how they transform or filter data.
Let \( F_{\text{single}} = B \) be a baseline single-agent model and \( F_{\text{multi}} = B \circ A \) be a two-agent pipeline. Then, under the same perturbation scenario (e.g., input or model-based), if
\[
\mathcal{M}(F_{\text{multi}}; \mathbb{P}) < \mathcal{M}(F_{\text{single}}; \mathbb{P}),
\]
we say that agent \( A \) \emph{amplifies} the vulnerability. If the inequality is reversed, \( A \) \emph{attenuates} the vulnerability. 

To explain this further, we introduce the concept of directional alignment in feature space. Let \( \Delta_A \) be the direction in which agent \( A \) transforms or maps data, and \( \Delta_B \) be the direction to which agent \( B \) is most sensitive (e.g., vulnerable). The alignment score \( \langle \Delta_A, \Delta_B \rangle \) indicates how likely \( A \)'s transformation pushes data toward \( B \)'s vulnerable zone. Positive alignment indicates amplification, while orthogonality or negative alignment suggests attenuation. The work by Wang et al.~\cite{wang2024demystifying} provides supporting evidence for this insight into the single-agent backdoor problem.
This insight generalizes to deeper pipelines: when all \( \Delta_i \) directions reinforce each other, vulnerability compounds; when directions interfere destructively, malicious signals may be filtered out.

\begin{figure}[htbp]
\centering
\includegraphics[width=0.9\textwidth]{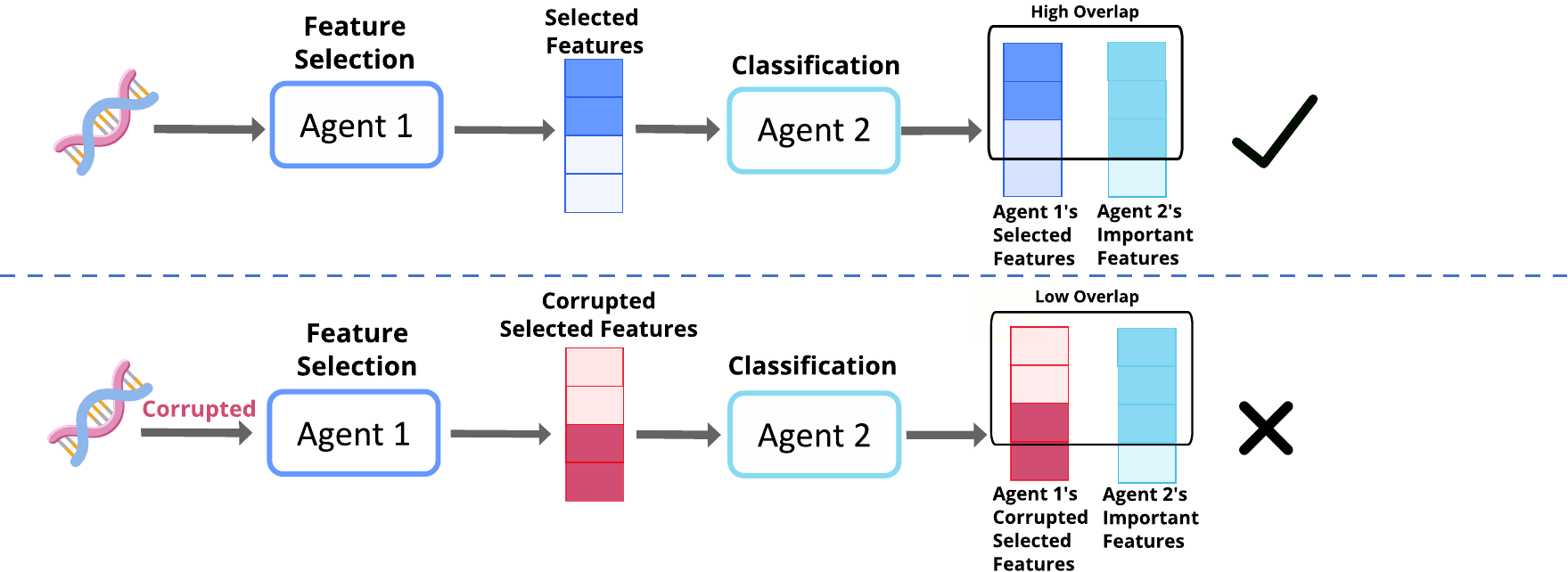}
\caption{Illustration of our experiment on propagation of attribution errors in a two-agent system for cell-type classification. \textbf{Top}: Clean RNA input is passed to Agent 1 for feature selection. Selected features produced by Agent 1 are passed to Agent 2 for classification. \textbf{Bottom}: Corrupted RNA input is passed to Agent 1 for feature selection. Corrupted selected features produced by Agent 1 magnify this corruption, and are passed to Agent 2 for classification. The corruption is further magnified in the final performance.}
\label{fig:vulnerability}
\end{figure}

\vspace{0.1cm}
\textbf{Empirical Simulation: Propagation of Attribution Errors in Two-Agent Pipelines.}
We design a hypothetical simulation inspired by multi-modal single-cell experiments~\cite{tasic2018shared}. In such biological settings, some cells are measured with both RNA expression and electrophysiological signals (Agent 1), while others only have RNA and cell-type labels (Agent 2). Due to biological relevance, genes important for predicting signal strength in Agent 1 are often also informative for predicting cell type in Agent 2. This motivates a multi-agent collaboration strategy in which Agent 1 performs feature selection on large-scale multi-modal data, and Agent 2 reuses the selected features for downstream classification. Figure~\ref{fig:vulnerability} illustrates this setup.

To test the robustness of this feature transfer, we compare three approaches across 10 independent simulations: \textbf{Clean RF:} Agent 1 selects features via Random Forest on clean data $(\mathbf{X}_1, y_{\text{signal}})$, which are then used by Agent 2 for cell-type classification. \textbf{Corrupted RF:} Agent 1’s input features are corrupted by weakening the signal-bearing genes (via scaling) and injecting Gaussian noise. Agent 2 uses the features selected from this corrupted dataset. \textbf{All Features:} Agent 2 uses all 1000 features without selection.
Agent 1 is trained on 1000 samples and Agent 2 on only 200, simulating a common high-dimensional, low-sample regime in single-cell analysis. The downstream performance (Macro-F1) of Agent 2 is summarized in table~\ref{tab:agent2_macro_f1}.

Although the regression output $y$ remains unchanged, corruption of $\mathbf{X}_1$ severely degrades Agent 1’s feature attribution. This leads to substantial overlap reduction and downstream performance drop. In this example, we illustrate that \textbf{multi-agent pipelines can silently propagate upstream errors}, resulting in worse performance than a single-agent baseline.

This case highlights a key risk in MAS pipelines: errors made by upstream agents can covertly bias downstream agents, even without adversarial intent. Unlike traditional backdoor attacks, these errors emerge from structural misalignment rather than explicit poisoning.

\begin{table}[tb]
\centering
\scalebox{0.85}{
\begin{tabular}{lccc}
\toprule
\textbf{Setting} & Clean RF & Corrupted RF & All Features \\
\midrule
\textbf{Mean Macro-F1} & \textbf{0.461} & \textbf{0.330} & \textbf{0.380} \\
\bottomrule
\end{tabular}
}
\vspace{0.1cm}
\caption{Mean Macro-F1 performance of Agent 2 under different feature sources.}
\label{tab:agent2_macro_f1}
\end{table}

\subsection{Impact of Topology on Backdoor Amplification and Attenuation}

The previous sections formalized how vulnerabilities in MAS, such as backdoor attacks, can propagate and either amplify or attenuate depending on directional alignment between agents. However, a key underexplored factor in this propagation is the \emph{topology} of the multi-agent AI system itself. That is, beyond the behavior between any two agents, the system’s overall graph structure can fundamentally shape how perturbations travel and interact.

Understanding the role of topology is crucial for two reasons. First, it governs the extent to which a single malicious signal can spread or get suppressed. Second, many real-world MAS operate under fixed or constrained topologies (e.g., star or cascade), making topological robustness a critical design constraint. Empirical studies such as~\cite{wang2024demystifying,li2021invisible,chen2017targeted} have shown that identical agent vulnerabilities can lead to vastly different outcomes depending on whether agents operate in centralized (star) or sequential (cascade) architectures. However, formal characterizations of how such topological differences result in varied outcomes are still limited. We illustrate these effects in the following using two canonical MAS topologies, star and cascade, and formalize their backdoor vulnerability propagation using performance metrics.

\paragraph{Star Topology.}
In a star topology, each agent \( g_i: \mathcal{X}_i \to \mathcal{Y}_i \) processes its input independently, and their outputs are aggregated by a central function \( h \). The full system is:
\[
F_{\mathrm{star}}(X_1,\dots,X_m) = h\bigl(g_1(X_1),\,g_2(X_2),\,\dots,\,g_m(X_m)\bigr).
\]
If multiple agents share similar backdoor vulnerabilities—such as reacting to a common trigger \(\nu_X\)—then their outputs may correlate, causing the aggregated output to reinforce the malicious effect. Formally, amplification occurs if:
\[
\mathcal{M}(F_{\mathrm{star}}; \mathbb{P}) \;<\; \max_{1 \le i \le m} \mathcal{M}(g_i; \mathbb{P}).
\]
This condition indicates that the MAS performs worse than its weakest agent under attack, due to compounded vulnerabilities.

Conversely, attenuation occurs if the agents are diverse in their vulnerabilities or the aggregator is robust (e.g., via voting or median):
\[
\mathcal{M}(F_{\mathrm{star}}; \mathbb{P}) \;>\; \max_{1 \le i \le m} \mathcal{M}(g_i; \mathbb{P}),
\]
meaning the system as a whole mitigates backdoor effects better than any individual agent.

\paragraph{Cascade Topology.}
In a cascade topology, each agent’s output feeds into the next, forming a pipeline:
\[
X_{i+1} = g_i(X_i), \quad \text{for } i = 1, \dots, m-1,
\]
with \( X_1 \sim (1-\alpha)\mu_X + \alpha\nu_X \). The full pipeline becomes:
\[
F_{\mathrm{cascade}}(X_1) = g_m \circ g_{m-1} \circ \dots \circ g_1(X_1).
\]
Cascade structures can amplify vulnerabilities when early-stage transformations align with later agents’ trigger directions (i.e., when \(\langle \Delta_i, \Delta_{i+1} \rangle > 0\) repeatedly). Amplification arises when:
\[
\mathcal{M}(F_{\mathrm{cascade}}; \mathbb{P}) \ll \mathcal{M}(g_1; \mathbb{P}),
\]
signifying severe degradation relative to the first agent’s performance.

On the other hand, attenuation emerges when intermediate agents disrupt the signal path, scrambling or filtering malicious directions. Formally:
\[
\mathcal{M}(F_{\mathrm{cascade}}; \mathbb{P}) > \mathcal{M}(g_1; \mathbb{P}),
\]
implying that sequential transformations dilute the backdoor effect.


\section{Conclusion}

This paper presents a formal and systematic framework for analyzing multi-agent AI systems (MAS), with a focus on effectiveness and safety. We establish mathematical definitions for agent interactions, dynamic topologies, and feedback mechanisms, and extend the classical MAS formulation to open-network settings via the Internet of MAS. 

On the effectiveness side, we identify three key factors: task allocation, robustness, and feedback. First, we show that MAS can outperform single-agent systems when tasks can be flexibly divided and reallocated among agents in response to real-time feedback. Second, redundancy among agents can also enhance robustness, but only when their training data is sufficiently diverse. Our experiments demonstrate that high overlap in training data undermines the benefit of ensembling by making agent decisions highly correlated. Third, we show that structured feedback, both from humans and between agents, enables MAS to adapt not only their own behaviors but also their topologies over time. 

On the safety side, we formalize how vulnerabilities, such as backdoor attacks, can propagate or amplify within MAS due to agent interdependence. Our analysis highlights that both the directional alignment between agents and the topology of the multi-agent system  influence how vulnerabilities propagate, either amplifying or attenuating errors across the system. Empirical simulations reveal that upstream errors or misaligned objectives can silently degrade downstream performance, sometimes making multi-agent pipelines more fragile than their single-agent counterparts. 

Overall, this work provides a structured foundation for understanding and evaluating MAS from the perspectives of effectiveness and safety. We hope this framework can inform and inspire future research toward building more capable and reliable multi-agent AI systems. Moving forward, several important directions remain to be explored. These include developing formal metrics for inter-agent influence and failure propagation, designing coordination protocols that are robust to agent misalignment, and ensuring the safe deployment of MAS in open, heterogeneous environments. In particular, maintaining trust and cooperation across organizational boundaries continues to be a significant challenge for real-world applications.

\bibliographystyle{unsrtnat}
\bibliography{references}

\end{document}